\documentclass[]{aastex701}

\usepackage{float}
\usepackage{placeins}
\usepackage{caption}
\usepackage{natbib}
\usepackage{xspace}
\begin{document}

\title{Optimizing Color Selection for Cluster-Member Identification with SuperBIT}

\newcommand\northeastern{Department of Physics, Northeastern University}
\newcommand\bls{Boston Latin School}
\newcommand\stanford{Stanford University}
\defcitealias{Saha:2026rbf}{LITB III}
\defcitealias{litb2}{LITB II}

\author[0009-0007-7736-2033]{Leila Ohashi}
\affiliation{\stanford}
\affiliation{\bls}
\affiliation{\northeastern} 
\email[]{lohashi@stanford.edu}  

\author[0000-0002-6044-2164]{Sayan Saha}
\affiliation{\northeastern}
\email{sa.saha@northeastern.edu}

\author[0000-0002-9883-7460]{Jacqueline E.\ McCleary}
\affiliation{\northeastern}
\email{j.mccleary@northeastern.edu}


\collaboration{all}{SuperBIT Collaboration}
\correspondingauthor{Leila Ohashi}
\email{lohashi@stanford.edu}

\begin{abstract}
Galaxy richness maps provide a useful tracer of the stellar mass distribution in galaxy clusters, but identifying cluster members from multiband photometric data can be challenging. In this work, we investigate multiband imaging data obtained with the Super-Pressure Balloon-borne Imaging Telescope (SuperBIT) during its science flight in April 2023. Using spectroscopically identified cluster members across the SuperBIT cluster sample, we benchmark the performance of the three colors available from the three SuperBIT bands: $(u-b)$, $(b-g)$, and $(u-g)$. We quantify the performance of each color using the $1\,\sigma$ scatter of the spectroscopically identified members in color and the purity of our cluster-member identification algorithm. We find that $(b-g)$ consistently exhibits the smallest color scatter and highest purity across the sample and over the redshift range probed by SuperBIT. We therefore identify $(b-g)$ as the preferred color for red-sequence-based cluster-member identification, providing a basis for future analyses of the cluster galaxy population using SuperBIT multiband photometry.

\end{abstract}


\section{Introduction} 
Galaxy clusters are the largest gravitationally bound structures in the Universe and provide important probes for constraining cosmic evolution of the Universe~\citep{intro1} and understanding the nature of dark matter~\citep{Clowe:2003tk}. Weak gravitational lensing offers a direct probe of the total projected mass of a cluster, which is predominantly dark matter, while the baryonic component can be traced using independent observables such as galaxy richness, X-ray emission, and the Sunyaev-Zeldovich (SZ) effect.

The Super-pressure Balloon-borne Imaging Telescope ~\citep[SuperBIT;][]{Gill2024} conducted its science flight in April 2023 and obtained imaging data for thirty merging galaxy clusters in three bands at the blue end of the optical spectrum: F400W ($u$), F480W ($b$), and F600W ($g$). The weak-lensing analysis pipeline is  described in detail in Lensing in the Blue II~\citep[LITB II hereafter]{litb2} and III~\citep[LITB III hereafter]{Saha:2026rbf}. To probe the baryonic component, and in particular the stellar mass, it is important to accurately identify galaxies that are gravitationally bound to the cluster. Cluster-member galaxies share common photometric properties that can be exploited for this purpose. Galaxies in clusters tend to have predominantly older stellar populations and relatively low star-formation rates. The accumulation of heavier elements through successive generations of star formation also produces a prominent $4000\,$\AA{} break in their spectra. This spectral feature gives rise to the well-known ``red sequence line'' in galaxy color--magnitude diagrams.

The overarching goal of this work is to improve the accuracy of the cluster-member identification algorithm implemented in the SuperBIT pipeline, which uses the red-sequence feature to identify likely cluster members. For the SuperBIT cluster sample, spanning redshifts of $0.1 \lesssim z \lesssim 0.5$, any of the three available colors, $(u-b)$, $(b-g)$, and $(u-g)$, could be used to identify the red sequence line in color--magnitude diagrams. In this work, we empirically quantify which of these colors most accurately captures the red-sequence feature.  In addition to the physical location of the 4000 \AA{} break, the accuracy with which the red sequence line can be constrained depends on the quality of the photometry in each band. In Section~\ref{sec:data}, we describe the SuperBIT photometric data and external data used in this analysis, along with a brief overview of our methods. In Section~\ref{sec:result}, we present and discuss our results.

\section{Data \& Methods} \label{sec:data}
\begin{figure*}[t]
    \centering
    \includegraphics[width=0.95\linewidth]{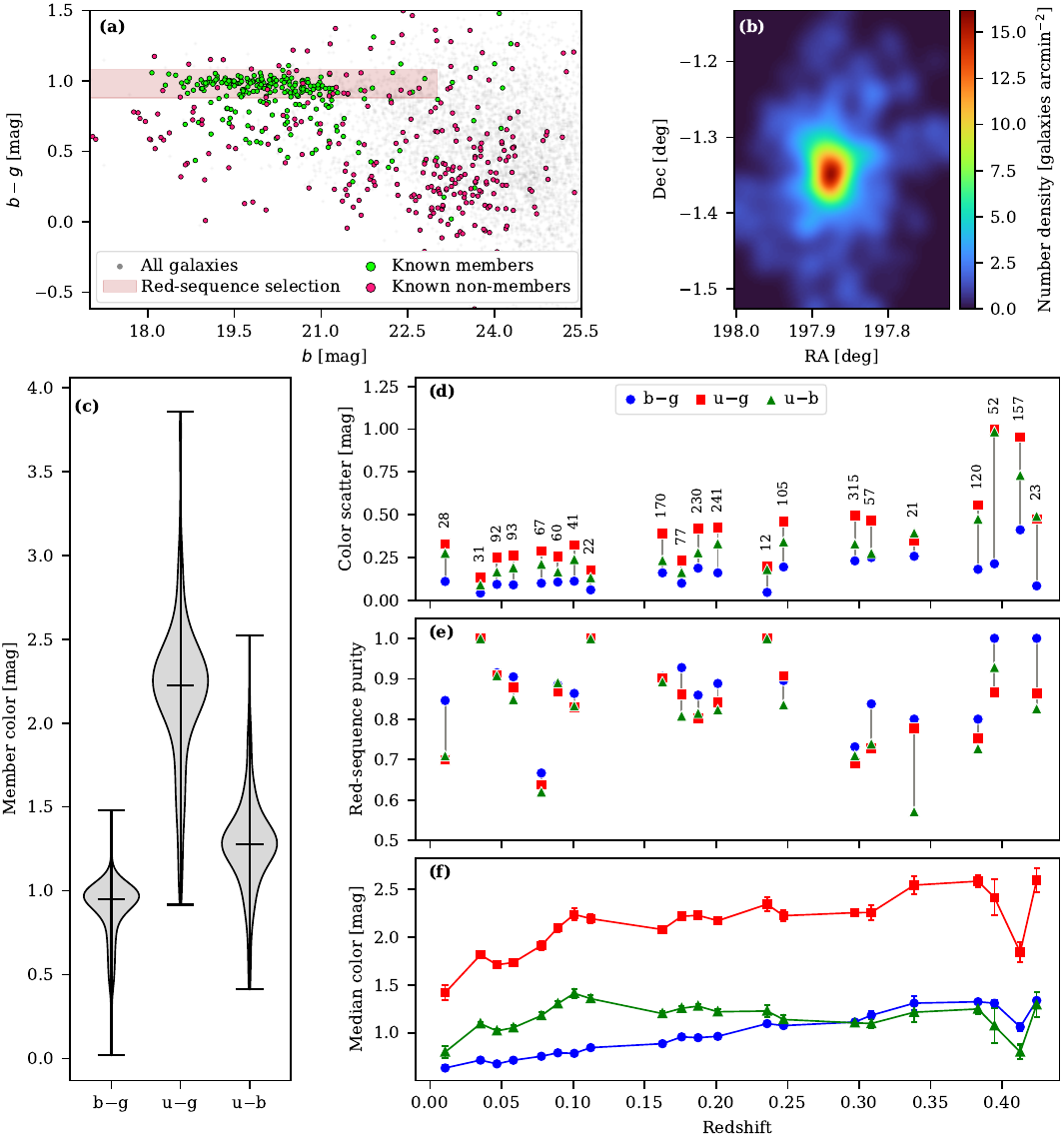}
\caption{
\emph{(a) :} Color--magnitude diagram using the $(b-g)$ color. Spectroscopically identified cluster members are marked with green circles, while all SuperBIT detections in the field of view are shown as gray points. The fitted red-sequence selection region is shown as a red box.
\emph{(b) :} The photometrically identified cluster members are mapped onto the sky in right ascension and declination to construct the projected galaxy number-density map of Abell 1689.
\emph{(c) :} Color distributions of the spectroscopically identified members of Abell 1689 for the three available colors, $(b-g)$, $(u-g)$, and $(u-b)$.
\emph{(d) :} The $1,\sigma$ color scatter of the spectroscopically identified members for all SuperBIT cluster targets as a function of cluster redshift. Each vertical grouping represents one target, with the number of available spectroscopically identified members annotated above the corresponding points. The scatter in $(b-g)$, $(u-g)$, and $(u-b)$ is shown by blue circles, red squares, and green triangles, respectively. The three targets with redshifts close to $z\sim0.1$ have been shifted slightly along the redshift axis for visual clarity.
\emph{(e) :} The red-sequence purity for the three available colors as a function of cluster redshift.
\emph{(f) :} The median color of the known members as a function of cluster redshift.
}
    \label{fig:result}
\end{figure*}
The full list of targets, comprising 30 galaxy clusters and five calibration fields for which photometry was obtained in all three SuperBIT bands, is given in Table 2 of \citetalias{Saha:2026rbf}. The photometric calibration is described in Section 4 of \citetalias{Saha:2026rbf}. The measured magnitudes were corrected for Galactic dust extinction using the extinction model of ~\cite{gordon2023} and the Galactic extinction map of ~\cite{Chiang:2023hyp}; details will be presented in \textit{Lensing in the Blue IV} (Saha et al., in preparation).
The extinction-corrected magnitudes are used to construct the three available colors, $(u-b)$, $(b-g)$, and $(u-g)$, which form the basis of our analysis.


The cluster-member identification procedure begins by selecting SuperBIT detections with external spectroscopic redshifts consistent with the redshift of the target cluster. Spectroscopic redshifts are obtained from the Dark Energy Spectroscopic Instrument ~\citep[DESI DR1;][]{DESI:2025fxa} and the NASA/IPAC Extragalactic Database (NED)\footnote{\url{https://ned.ipac.caltech.edu/}}~\citep{ned}.
We classify a matched object as a spectroscopically identified cluster member when its external redshift satisfies $|z_{\rm spec}-z_{\rm cluster}| < 0.02$. These spectroscopically identified members are then located in the color--magnitude diagram of each cluster. An example is shown by the green circles in the panel (a) of Figure~\ref{fig:result}. We characterize the red sequence line by fitting a linear relation to the color--magnitude data of the spectroscopically identified members. Candidate photometric members are subsequently selected within a color window of width 0.1 on either side of the best-fit red-sequence line. We additionally impose an upper magnitude limit that depends on the cluster redshift, which reduces contamination from faint background objects. 

The accuracy of this photometric cluster-member selection is dependent, in part, on the precision of the colors of the spectroscopically identified members used to determine the red sequence. In this work, we quantify this precision by measuring the $1\, \sigma$ color scatter of the spectroscopically identified members for each of the three available colors, $(u-b)$, $(b-g)$, and $(u-g)$. 

\section{Results and Discussion} \label{sec:result}

As an example of the cluster-member identification algorithm implemented in the SuperBIT pipeline, we show the results for Abell 1689 in panel (a, b) of Figure~\ref{fig:result}. In panel (a), we show the color--magnitude diagram with $(b-g)$ color and $b$ magnitude, with all SuperBIT detections in the field shown as gray points and the spectroscopically identified cluster members described in Section~\ref{sec:data} shown as green circles. Spectroscopically identified non-members, which do not fall within the redshift slice around the cluster redshift, are shown as red circles. The red box denotes the region around the best-fit red-sequence relation used to select photometric cluster-member candidates. Objects falling within this selection region are then mapped into right ascension--declination space to construct the corresponding projected member-galaxy number-density map, shown in panel (b).

In panel (c) of Figure~\ref{fig:result}, we compare the distributions of the three available colors, $(u-b)$, $(b-g)$, and $(u-g)$, for the spectroscopically identified cluster members of Abell 1689 using violin plots. The $(b-g)$ color exhibits the narrowest distribution, indicating that it provides the smallest color scatter among the three available color combinations for the identified members. The quantitative comparison across the full SuperBIT cluster sample is summarized in panels (d) and (e). In panel (d), we show the $1\,\sigma$ scatter of the color distribution of the spectroscopically identified members as a function of cluster redshift. Each vertical grouping of three points corresponds to one cluster, with the three points representing the scatter in $(b-g)$ (blue circles), $(u-b)$ (green triangles), and $(u-g)$ (red squares). The number of spectroscopically identified members used for each cluster is indicated adjacent to each grouping. In panel (e), we present the purity of the red-sequence catcher for the three available color combinations. We define the purity as the fraction of objects with known redshifts that are identified as cluster members by the red-sequence selection, i.e.,
\begin{equation}
    P = \frac{N_{\rm member}}{N_{\rm member}+N_{\rm non-member}},
\end{equation}
where $N_{\rm member}$ is the number of spectroscopically identified cluster members that fall within the red-sequence selection region, and $N_{\rm non-member}$ is the number of spectroscopically identified non-members that also fall within the same red-sequence selection region. These objects correspond to the green and red circles, respectively, inside the  red box in panel (a). Finally, in panel (f), we also show the evolution of the median color of the known cluster members as a function of redshift.

Across the full sample and over the range of cluster redshifts probed by SuperBIT, the $(b-g)$ color consistently exhibits the smallest scatter of the three color combinations and correspondingly tends to yield the highest purity. This result demonstrates that $(b-g)$ provides the most precise color measurement of the red-sequence population in the SuperBIT data, making it the preferred choice for the cluster-member identification algorithm.

The relative performance of the three colors depends not only on the spectral information captured by the corresponding filters, but also on the quality of the photometry in each band. In particular, the filter transmission efficiency (Figure~2 of \citetalias{litb2}) and sky background (Section~4.3 of \citealt{ajaythesis}) contribute to the photometric uncertainties. Among the SuperBIT bands, F480W ($b$) provides the best combination of transmission efficiency and sky-background performance, followed by F600W ($g$). Consequently, the $(b-g)$ color has the smallest photometric scatter among the three color combinations considered.

We therefore adopt $(b-g)$ as the primary color for photometric cluster-member identification in \textit{Lensing in the Blue IV}, using it in the color--magnitude selection to construct cluster-member number-density maps. These maps allow us to compare the spatial distribution of the stellar population traced by red-sequence galaxies with the weak-lensing signal measured from the SuperBIT observations. The red-sequence number-density maps will also provide an input for future coherence analyses~\citep{Cerini:2022akj, Cerini:2026jpm} using SuperBIT data.

\begin{acknowledgments}
The authors acknowledge Maya Amit and Marcus Michaud for their contributions to the photometry pipeline, as well as the entire SuperBIT collaboration. The authors also acknowledge valuable feedback from Eric Habjan and William C. Jones on the draft of the paper.
This work was completed using the Explorer Cluster, supported by Northeastern University's Research Computing team. This research has made use of the NASA/IPAC Extragalactic Database, which is funded by the National Aeronautics and Space Administration and operated by the California Institute of Technology.
\end{acknowledgments}


\bibliography{main}{}
\bibliographystyle{aasjournalv7}



\end{document}